\documentclass{JHEP3}

\usepackage{epsfig} 

\def\be{\begin{equation}}
\def\ee{\end{equation}}
\def\bear{\begin{eqnarray}}
\def\eear{\end{eqnarray}}
\def\nn{\nonumber}

\def\half{{{1\over 2}}}

\newcommand\inv[1]{{1\over{#1}}}

\newcommand\Tr[1]{{\mbox{Tr}{#1}}}          

\def\a{{\alpha}}

\def\s{{\sigma}}

\newcommand{\ds}{\frac{d}{d\sigma}}

\author{ Joanna L. Karczmarek \\
Jefferson Physical Laboratory, Harvard University, Cambridge, MA, 02138
\email{karczmarek@physics.harvard.edu}}

\title{Multicore Noncommutative Bions}

\abstract{ The noncommutative bion core of Constable, Myers and
Tafjord \cite{Constable:1999ac} describes the BPS D1-D3 brane
intersection (where a single bundle of D1-branes is attached to the
D3 brane) in the nonabelian Born-Infeld theory of D1-branes. The
possibility of extending this construction to BPS configurations
with multiple separated parallel bundles of D1-branes attached to a
{\it single} D3-brane is discussed. The problem is reduced to
solving the Nahm equation with novel boundary conditions. A
concrete, non-trivial solution is presented.}

\begin{document}

\section{Introduction}

Among the more interesting properties of D-branes are the many
relationships between D-branes of various dimensions. For example,
lower dimensional branes can be described as solitons in the
effective theory of a higher dimensional D-brane
\cite{Douglas:1995bn}; they are then seen as bound to the higher
dimensional brane. Wrapping a brane around a thin tube supported the
appropriate flux produces an object effectively identical to that of
a lower dimensional brane \cite{Emparan:1997rt}. In similar spirit,
brane intersections between lower and higher dimensional branes can
be described in the effective theory of the higher dimensional
brane: the brane develops a spike which is interpreted as an
attached bundle of lower dimensional branes.  For example, for a
D3-brane, one finds that a magnetic monopole  of strength $N$
produces a singularity in the D3-brane's transverse displacement:
this is the  spike which can be interpreted as $N$ D1-branes
attached to the D3-brane \cite{Callan:1997kz, Gibbons:1997xz}.

With the development of nonabelian effective actions
for stacks of D-branes \cite{Tseytlin:1997cs,Myers:1999ps}
\footnote{See references in these papers for other work on this problem.},
it became possible to describe higher dimensional objects as made
up of lower dimensional components.  As is well known, the gauge group
for a stack of $N$  superposed D-branes is enhanced
from $U(1)^N$ to $U(N)$ and  the brane worldvolume supports a $U(N)$
gauge field as well as a set of scalars in the adjoint representation
of $U(N)$ (one for each of the transverse coordinates).
A striking feature of this effective action is that it contains
structures such as commutators of the transverse coordinates
with themselves. These vanish in the $U(1)$ case and cannot be directly
inferred from the abelian Born-Infeld action.
They lead to nonabelian geometrical structures, where higher
dimensional branes can be built from lower dimensional branes,
via the dielectric effect \cite{Myers:1999ps}, or its analogs.

In \cite{Constable:1999ac}, the nonabelian Born-Infeld
action was applied to a stack of $N$ D1-branes
in a flat background and it was shown that the transverse coordinates
of the D1-branes `flare out' to a flat three dimensional space.
This was interpreted as a collection of D1-branes attached to
an orthogonal D3-brane, one of the standard D-brane intersections
mentioned above.  This then was the dual description of the
point magnetic charge in the effective theory of a
single (abelian) D3-brane.  To the extent that it is possible to compare them,
there is a perfect quantitative match between the two pictures.

The D1-D3-brane intersection is supersymmetric, which is reflected
in both pictures, since the nontrivial solutions describing the
intersection are BPS.  In case of the nonabelian Born-Infeld action
for the D1-branes, the BPS equation was motivated in
\cite{Constable:1999ac} to be the Nahm equation.  This is quite
natural, since the Nahm equation has made its appearance in similar
contexts before (see section \ref{sec:comments}).

Various other brane intersections have also
been described this way \cite{Constable:2001ag,Constable:2001kv,
Karczmarek:2001pn,Constable:2002yn,Cook:2003rx}.

In this paper, we will try to construct multiple separated parallel
bundles of D1-branes flaring out to form a {\it single} D3-brane.
We will argue that the existence of such configurations is obvious
from the dual abelian description on a D3-brane. The construction in
the nonabelian theory of D1-branes is nontrivial, and will require
finding a solution to the Nahm equation with novel boundary
conditions. We will explicitly present one such solution in section
\ref{sec:existence}.

\section{Dual pictures for the D1-D3 brane intersection: review}
\label{sec:review}

Consider the abelian Born-Infeld action for a single
D3-brane in flat space. Let the D3-brane extend in 0123-directions,
and let the coordinates on the brane be denoted by $x^i$,
$i=0,\ldots,3$. Restricting the brane to have displacement in
only one of the transverse directions, we can take the (static gauge)
embedding coordinates of the brane in the ten-dimensional space to be
$X^i = x^i$, $i=0,\ldots,3$; $X^a = 0$, $a=4,\ldots,8$;
$X^9 = \sigma(x^i)$. Then there exists a static (BPS) solution of
the Born-Infeld action corresponding to placing $N$ units of
$U(1)$ magnetic charge at the origin of coordinates on the brane
\cite{Callan:1997kz}:
\be
X^9(x^i) = \sigma(x^i) = {{q} \over {\sqrt{(x^1)^2+(x^2)^2+(x^3)^2}}} ~,
\label{eqn:bion}
\ee
where $q=\pi\alpha' N$, and $N$ is an integer. This magnetic
bion solution to the Born--Infeld action corresponds to $N$ superposed
D1-branes attached to the D3-brane at the origin. It is ``reliable''
in the sense that the effect of unknown higher-order corrections in
$\alpha'$ and $g$ to the action can be made systematically small
in the large-$N$ limit (see \cite{Callan:1997kz} for details).
At a fixed $X^9=\sigma$, the cross-section of the deformed
D3-brane is a 2-sphere with a radius
\be
r(\sigma) = {{ \pi\alpha' N} \over {\sigma}}~.
\label{eqn:r0}
\ee
From the point of view of the D3-brane, magnetic monopole formed
this way is a BPS object and there is no net force between two of
these objects placed at a finite distance.  In fact, a more general
solution is given simply by a sum of $k$ such spikes
\be
X^9(x^i) = \sigma(x^i) =
\sum_{a=1}^{k}~
{q_{(a)} \over {\sqrt{(x^1-x^1_{(a)})^2+(x^2-x^2_{(a)})^2+(x^3-x^3_{(a)})^2}}} ~,
\label{eqn:bions}
\ee
each of which has charge $q_{(a)}$ and position $(x^1_{(a)},x^2_{(a)},x^3_{(a)})$.

Having described the D1-D3 intersection from the point of view of the
D3-brane, we now move onto the dual description in terms of the D1-brane
action \cite{Constable:1999ac}.

We consider the nonabelian Born--Infeld action 
specialized to the case of N coincident D1-branes,
in flat background spacetime, with vanishing
$B$ field, vanishing worldvolume gauge field and constant dilaton.
The action depends only on the $N\times N$ matrix transverse
scalar fields $\Phi^i$'s. In general, $i=1,\ldots,8$, but since we are
interested in studying the D1/D3-brane intersection, we will allow
only three transverse coordinate fields to be active ($i=1,2,3$).
The explicit reduction of the static gauge action ($X^0 = \tau$ and
$X^9=\sigma$) is then \cite{Tseytlin:1997cs}
\be
\label{eqn:SBI1}
S_{BI} = -T_1 \int d\sigma d\tau STr
 \sqrt{ -\det(\eta_{ab}+(2\pi\a')^2\partial_a\Phi^i
   Q^{-1}_{ij}\partial_b\Phi^j) det(Q^{ij}) } ~,
\ee
where
\be
Q^{ij} = \delta^{ij} + i 2\pi\a' [\Phi^i,\Phi^j]~.
\ee
Since the dilaton is constant, we incorporate it in the
tension $T_1$ as a factor of $g^{-1}$.

The solution we are interested in is a static, BPS solution.
It can be argued \cite{Constable:1999ac} that the BPS condition is
the Nahm equation,
\be
\partial \Phi_i =  \half i \epsilon_{ijk}[\Phi^j,\Phi^k] ~.
\label{eqn:BPS}
\ee

The trivial solution has  $\Phi^{ij}=0$, which
corresponds to an infinitely long bundle of coincident D1-branes.
In \cite{Constable:1999ac}, a much more interesting solution was found
by starting with the following ansatz:
\be
\Phi^i = 2 \hat R(\sigma) \alpha^i,\qquad\qquad
(\alpha^1, \alpha^2, \alpha^3) \equiv {\bf{X}} ~,
\label{eqn:ansatz0}
\ee
where $\alpha^i$ are the $N\times N$-dimensional
generators of a representation of the $SU(2)$ subgroup of $U(N)$,
$[\alpha^i, \alpha^j] = i\epsilon_{ijk}\alpha^k$.
When the ansatz is substituted into the BPS condition
(\ref{eqn:BPS}), we obtain a simple equation for $\hat R$,
\be
\hat R'= - 2\hat R^2 ~,
\ee
which is solved by
\be
\hat R =  \inv{2 \sigma}~.
\label{eqn:R}
\ee

This solution maps very nicely onto the bion solution of the
previous section.  At a fixed point $|\sigma|$ on the D1-brane stack,
the geometry given by (\ref{eqn:ansatz0}) is that of a sphere
with the physical radius $R^2={(2\pi\a')^2\over N}Tr (\Phi^i)^2$.
For the ansatz under consideration, this gives
\be
R(\sigma)^2 = {{(2\pi\a')^2}\over {N}} Tr (\Phi^i)^2
= (2\pi\a')^2 \hat R(\sigma)^2 C ~,
\ee
where $C$ is the quadratic Casimir, equal to $N^2-1$
for an irreducible representation of $SU(2)$. Therefore,
\be
R(\sigma)= {{\pi\a' \sqrt{N^2-1}} \over {(|\sigma|)}}
\cong {{\pi\alpha' N} \over {|\sigma|}}
\ee
for large N, in agreement with equation (\ref{eqn:r0}). This
completes our synopsis of the arguments given in \cite{Constable:1999ac} for
the agreement between the commutative and noncommutative approaches
to the D1/D3-brane intersection.

The natural question to ask is how to obtain an analog of the
multi-bion solutions (\ref{eqn:bions}) in the D1-brane description.
This would require finding a solution to the BPS equation
(\ref{eqn:BPS}) with the following properties
\footnote{We are proposing these boundary conditions based
on geometrical considerations alone, but it should be possible
to derive them in a way similar to \cite{Chen:2002vb}.}. 
At $\s \rightarrow
\infty$, we would like a number of parallel D1-brane bunches,
separated in space; therefore the three matrices $\Phi^i$ should be
of the form \be \label{eqn:spikes} \Phi^i(\sigma \rightarrow \infty)
\sim {\mbox{diag}} \left( x^i_{(1)} {\bf I}_{q_{(1)}\times q_{(1)}}
+ 2 \hat R ~\alpha^i_{q_{(1)}}, ~~\dots,~~ x^i_{(k)} {\bf
I}_{q_{(k)}\times q_{(k)}} + 2 \hat R ~\alpha^i_{q_{(k)}} \right )~,
\label{eqn:multicore} \ee where $\alpha^i_{q_{(a)}}$ are
$q_{(a)}$-dimensional generators of an irreducible representation of
SU(2). At $s=0$, though, the residue must be an  {\it irreducible}
representation of SU(2), so that a {\it single} D3-brane is formed:
\be \label{eqn:brane} \Phi^i(\sigma \rightarrow 0) \sim 2 \hat R
~\alpha^i_{N} ~+~{\textrm{finite}}, \label{eqn:multicore0} \ee

The desired solution  to (\ref{eqn:BPS}) must then
interpolate between $k$ irreducible representations
of SU(2) at large $\s$, and the single irreducible representation
at small $\s$.

Duality between the two descriptions of this BPS brane intersection
strongly suggests that such solutions exists for arbitrary
$k$, $q_{(a)}$s and $x^i_{(a)}$s.  In the next section we will
solve a simple but nontrivial example with $k$=2 and
$q_{(1)}=q_{(2)}=2$.

\section{Multicore example}
\label{sec:existence}

We begin with the following well-known observation:
the Nahm equation can be rewritten in Lax form, by defining
\footnote{This is a special case, sufficient for out purpose;
for a more general case,
see for example a review in \cite{Sutcliffe:1997ec}.}
\be
M \equiv \half \left ( \Phi^1 + i \Phi^2 \right )
~,~~ L \equiv  \half \Phi^3~.
\ee
The Nahm equation implies that
\be
\ds M =  [M ,L]~,
\label{eqn:lax}
\ee
and therefore $\Tr (M^k)$ is a constant for all k.
These constants of motion will be useful in our analysis.

Consider the following ansatz for a solution to equation
(\ref{eqn:BPS})
\bear
\Phi^1(\sigma) &=& 2\left [
\begin{array}{cccc}
-A(\sigma) & 0 & 0 & 0 \cr
0 & -B(\sigma) & 0 & 0 \cr
0 & 0 & B(\sigma) & 0 \cr
0 & 0 & 0 & A(\sigma)
\end{array}
\right ]~,
\nn \\ ~
\nn \\
\Phi^2(\sigma) &=& 2\left [
\begin{array}{cccc}
0 & C(\sigma) & 0 & 0 \cr
C(\sigma) & 0 & D(\sigma) & 0 \cr
0 & D(\sigma) & 0 & C(\sigma) \cr
0 & 0 & C(\sigma) & 0
\end{array}
\right ]~,
\\ \nn ~
\\ \nn
\Phi^3(\sigma) &=& 2\left [
\begin{array}{cccc}
0 & iC(\sigma) & 0 & 0 \cr
-iC(\sigma) & 0 & iD(\sigma) & 0 \cr
0 & -iD(\sigma) & 0 & iC(\sigma) \cr
0 & 0 & -iC(\sigma) & 0
\end{array}
\right ]~.
\eear
For this ansatz, the Nahm equation (\ref{eqn:BPS}) reduces to
\be
\label{eqn:ABCD}
\left \lbrace \begin{array}{rcl}
\ds A &=& - 2C^2 \cr
\ds B &=& 2C^2 - 2D^2 \cr
\ds C &=& (B-A) C\cr
\ds D &=& - 2 BD\cr
\end{array} \right .
\ee
For future reference, let us write down a few solutions to this system
of equations.  The first is a block diagonal solution
given by putting together two $2 \times 2$ irreps of SU(2), separated by
an arbitrary distance $2 \Delta$
\be
\label{eqn:soln22}
A = \Delta + \hat R(\sigma)~,~~B = \Delta - \hat R(\sigma)~,
~~C = \hat R(\sigma)~,~~D = 0~.
\ee
The second solution corresponds to a single $2 \times 2$ irrep of SU(2),
together with some nonzero diagonal elements
\be
\label{eqn:soln2}
A = \Delta~,~~B = \Delta + \hat R(\sigma)~,~~C = 0~,~~D = \hat R(\sigma)~.
\ee
The third solution is the $4 \times 4$ irrep of SU(2)
\be
\label{eqn:soln44}
A = 3 \hat R(\sigma)~,~~B = \hat R(\sigma)~,~~C = \sqrt{3} \hat R(\sigma)~,~~D = 2 \hat R(\sigma)~.
\ee
The goal of this section is to prove that there exists a solution
which interpolates between solution (\ref{eqn:soln44}) for
$\sigma \rightarrow 0$ and (\ref{eqn:soln22}) for
$\s \rightarrow \infty$ according to equations
(\ref{eqn:spikes}) and (\ref{eqn:brane}).
  Such a solution should be interpreted
as describing two bions, each made up of 2 D1-branes, parallel
and separated by $4\Delta$, merging to funnel out into a single
D3-brane at $\s=0$.  The solution needs to extrapolate between
one noncommutative sphere with $N=4$ and two noncommutative spheres
with $N=2$ each.

Equation (\ref{eqn:lax}) leads to two convenient constants of motion
$c_1$ and $c_2$ defined by \bear c_1 &\equiv& \half Tr { (M^2)} =
A^2 + B^2 - 2C^2 - D^2~, \\ \nn c_2 &\equiv& - \half\left(\half
\Tr{(M^4)} - \left ( \Tr{(M^2)}\right)^2 \right) = (C^2 + AB)^2 -
A^2 D^2~. \eear For the $\s \rightarrow \infty$ asymptotic solution
of interest, (\ref{eqn:soln22}), these constants take values \be
\label{eqn:constants} c_1 = 2 \Delta^2~,~~c_2 = \Delta^4~. \ee
Solving (\ref{eqn:ABCD}) for $C$ and $D$, \be C^2= - \half \ds A~,~~
D^2 = -\half \ds (A + B) \ee and substituting into
(\ref{eqn:constants}), we obtain \bear A^2 + B^2 + \frac{3}{2}\ds A
+ \frac{1}{2} \ds B &=& 2 \Delta^2
\\ \nn
\left ( -\half \ds A + AB\right )^2 + \half A^2 \left(\ds A + \ds B
\right ) &=&  \Delta^4 \label{eqn:AB} \eear Finally, solving for
$\ds A$ and $\ds B$, \bear \ds A &=& 2 \left ( A^2 + AB - \sqrt{(A^2
+ AB)^2 + (A^2 - \Delta^2)^2} \right )~, \\ \nn \ds B &=& 2 \left (
-4A^2 - 3AB - B^2 + 2\Delta^2 + 3 \sqrt{(A^2 + AB)^2 + (A^2 -
\Delta^2)^2} \right )~. \label{eqn:ABsolved} \eear The sign on the
square root has been chosen to be consistent with the desired
asymptotic behavior.

\FIGURE{\epsfig{file=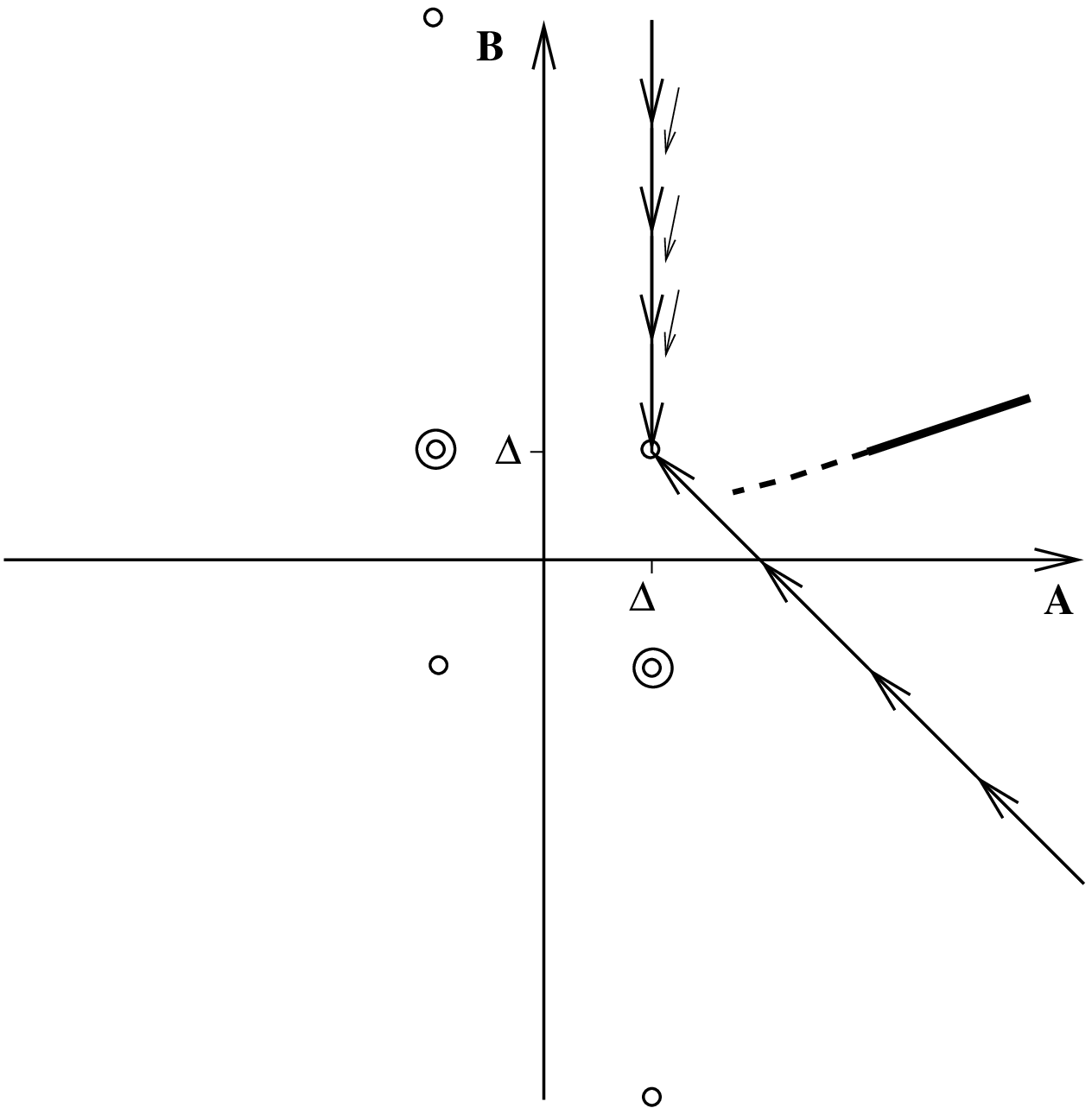} \caption{Solution flow in the A-B
plane. There are six critical points, two of which are double zeros.
Exact solutions (\ref{eqn:soln22}) and (\ref{eqn:soln2}) are shown
flowing to the critical point at $A=B=\Delta$.  The asymptotic
behavior of the solution of interest is also shown. The small arrows
indicate direction of the flow in the vicinity of $A=\Delta$ line. }
\label{fig:AB}}

The problem has now been reduced to first order equations for a flow
in the $A$-$B$ plane.  The two solutions (\ref{eqn:soln22}) and
(\ref{eqn:soln2}) correspond to $A+B = 2 \Delta$ and $A = \Delta$
respectively, and the direction of flow for $\s: 0 \rightarrow
\infty$ is show in Figure \ref{fig:AB}. The six critical points of
this flow, $\ds A = \ds B = 0$ are given by $\pm(A,B) = (\Delta,
\Delta),(\Delta, -\Delta), (\Delta, -5\Delta)$, and are also marked
in Figure \ref{fig:AB}. The interpolating solution we are interested
in is $A = 3B$ for  $\s\rightarrow 0$.  The question is whether this
solution approaches the critical point at $(A,B) = (\Delta, \Delta)$
as $\s\rightarrow\infty$. Uniqness implies that different solutions
cannot cross -- thus the solution we are interested in must stay
within the region bounded by lines $A+B = 2 \Delta$ and $A =
\Delta$.  In addition, $\ds A < 0 $ in that region for $A>\Delta$,
hence the solution must approach the $A=\Delta$ line for large $s$.
Finally, $\ds B < 0$ in the vicinity of the $A=\Delta$ line, hence
the flow cannot escape to infinity.  Therefore, the solution has
nowhere to `go' but the critical point at $(A,B) = (\Delta,
\Delta)$.


Once we know that our solution does in fact end at this critical
point, we need to analyze the local behavior around the critical
point to see whether the large $s$ behavior approaches the
two-bion-spike solution (\ref{eqn:soln22}). Expanding around the
critical point $A = \Delta + a$ and $B = \Delta + b$, we obtain to
lowest nontrivial order \bear
\ds a &=& - 2 a^2 ~,\\
\ds b &=& -4(a+b) + 4a^2 - 2b^2~. \eear We easily see that all
solutions within the region of interest approach the critical point
along the $b=-a$ line (along the zero eigenvalue direction), which
is precisely the solution we are interested in.

This completes the proof that there exists a solution to the BPS
equations which interpolates between the two-bion-spikes solution in
equation (\ref{eqn:soln22}) and the one-bion-spike solution in
(\ref{eqn:soln44}), corresponding to an interpolation between the
geometry of two noncommutative spheres with $N=2$ each and one
noncommutative sphere with $N=4$.  Figure \ref{fig:soln} shows the
result of a numerical computation of this interpolating solution in
the $A$-$B$ plane.

\FIGURE{\epsfig{file=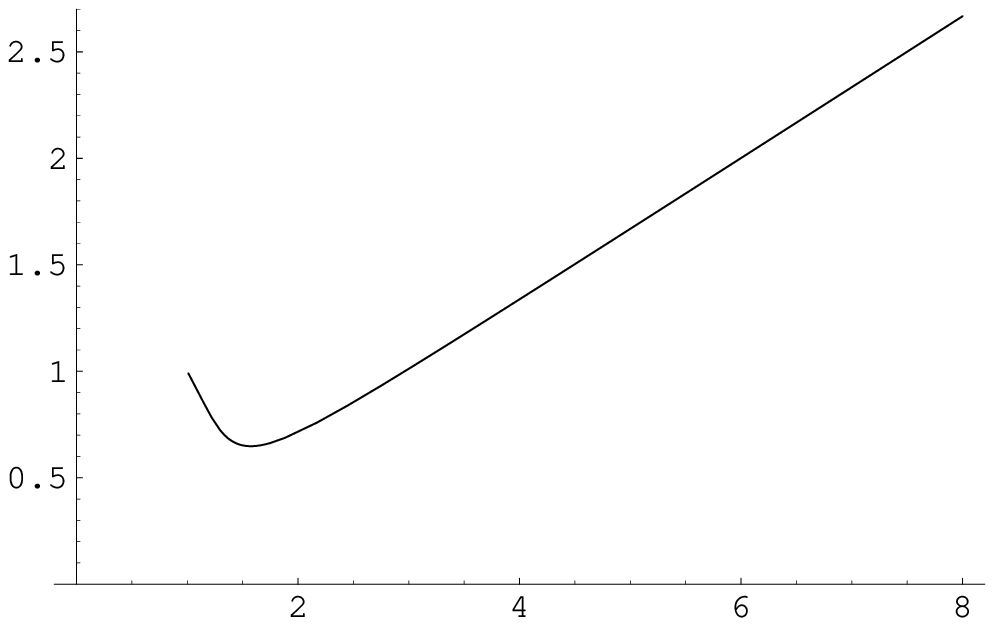} \caption{Interpolating solution in
the A-B plane, with $\Delta = 1$.  The solution is approximately
$B=A/3$ for large $A,B$ and approaches the $(1,1)$ point on the
$B=2-A$ line.} \label{fig:soln}}

Finally, lets mention that from equations
(\ref{eqn:AB}) it is a simple matter to obtain
an expansion for the solution around $\s = 0$:
\bear
A(\sigma) &=& \frac{3}{2\s} + \frac{2}{5} \Delta\s
- \frac{8}{175} (\Delta\s)^3 + o(s^5)
\\
B(\s) &=& \frac{1}{2\s} + \frac{2}{15} \Delta\s
+ \frac{104}{1575} (\Delta\s)^3 + o(s^5)
\eear

\section{Comments}
\label{sec:comments}

The appearance of the Nahm equation as a BPS condition for D1-branes
is quite natural and provides an interpretation of the Nahm
procedure in terms of lower dimensional branes, as was first pointed
out in \cite{Diaconescu:1996rk}. The Nahm equation arises in many
contexts and is solved with different boundary conditions, dictated
by the problem at hand. The standard boundary conditions  are those
which are useful in the ADHMN construction of monopoles. There, $\s$
is taken on the interval $(-1,1)$ and the matrices $\Phi^i(s)$ have
poles at $\s = \pm 1$ whose residues are generators of the same
irreducible representation of SU(2).  This corresponds to a bundle
of D1-branes connecting two parallel D3-branes separated by a finite
distance. By removing one of these poles to infinity so that $\s$
lives on the interval $(0,\infty)$, we remove one of the two branes
to infinity, and obtain the single D3-brane scenario described in
section \ref{sec:review}. More complicated boundary conditions,
describing the discontinuity as a bundle of D1-branes crosses a
D3-brane were discussed in 
\cite{Tsimpis:1998zh,Kapustin:1998pb,Chen:2002vb}.

The problem of solving the Nahm equations with
{\it different} representations at $\s=0$ and $\s=\infty$
was considered in \cite{Bachas:2000dx}.
There the boundary conditions
(\ref{eqn:multicore}) and (\ref{eqn:multicore0})
with all $x^i_{(a)}=0$ were considered.
In the geometrical language of this paper, there was no
separation between the individual D1-brane bundles ($\Delta=0$).
Dimensions of the moduli space of solutions were computed
(for example, for $2 + 2 \rightarrow 4$ and $\Delta=0$ the
moduli space is 4-dimensional).

Separating D1-branes ending on the same D3-brane was also considered
in \cite{Chen:2002vb}, where the boundary condition for removing
one D1-brane from the bundle was considered.

The solution presented here is a special case of a new boundary condition
for the Nahm equation,
given by equations (\ref{eqn:multicore}) and (\ref{eqn:multicore0}).
It would be interesting to see whether a more general analysis of existence
of solutions with such boundary conditions is possible.  Based on the
proposed duality between BPS objects in Born-Infeld theory of D3-branes
and D1-branes, such solutions to the Nahm equation should exist.

Unfortunately, $N=4$ is too small to obtain a three dimensional
geometry from our nonabelian solution.  It would be quite
interesting to check whether the nonabelian D1-brane picture
reproduces the correct shape of the deformed D3-brane, as given by
(\ref{eqn:bions}), but such geometric interpretation requires $N \gg
1$.

\section*{Acknowledgments}

I am grateful to Curt Callan and Rob Myers for helpful
discussions. I would also like to thank the Aspen Center for Physics
and University of Washington Physics Department for hospitality
while parts of this work were completed.
This work was supported in part by DOE grant DE-FG02-91ER40654.


\def\np#1#2#3{{\it Nucl.\ Phys.} {\bf B#1} (#2) #3}
\def\pl#1#2#3{{\it Phys.\ Lett.} {\bf B#1} (#2) #3}
\def\physrev#1#2#3{{\it Phys.\ Rev.\ Lett.} {\bf #1} (#2) #3}
\def\prd#1#2#3{{\it Phys.\ Rev.} {\bf D#1} (#2) #3}
\def\ap#1#2#3{{\it Ann.\ Phys.} {\bf #1} (#2) #3}
\def\ppt#1#2#3{{\it Phys.\ Rep.} {\bf #1} (#2) #3}
\def\rmp#1#2#3{{\it Rev.\ Mod.\ Phys.} {\bf #1} (#2) #3}
\def\cmp#1#2#3{{\it Comm.\ Math.\ Phys.} {\bf #1} (#2) #3}
\def\mpla#1#2#3{{\it Mod.\ Phys.\ Lett.} {\bf #1} (#2) #3}
\def\jhep#1#2#3{{\it JHEP} {\bf #1} (#2) #3}
\def\atmp#1#2#3{{\it Adv.\ Theor.\ Math.\ Phys.} {\bf #1} (#2) #3}
\def\jgp#1#2#3{{\it J.\ Geom.\ Phys.} {\bf #1} (#2) #3}
\def\cqg#1#2#3{{\it Class.\ Quant.\ Grav.} {\bf #1} (#2) #3}

\def\hepth#1{{\it hep-th/{#1}}}





\begin{thebibliography}{99}


\bibitem{Constable:1999ac}
N.~R.~Constable, R.~C.~Myers and O.~Tafjord,
Phys.\ Rev.\ D {\bf 61}, 106009 (2000)
[arXiv:hep-th/9911136].


\bibitem{Douglas:1995bn}
M.~R.~Douglas,
arXiv:hep-th/9512077.

\bibitem{Emparan:1997rt}
R.~Emparan,
Phys.\ Lett.\ B {\bf 423}, 71 (1998)
[arXiv:hep-th/9711106].



\bibitem{Callan:1997kz}
C.~G.~.~Callan and J.~M.~Maldacena,
Nucl.\ Phys.\ B {\bf 513}, 198 (1998)
[arXiv:hep-th/9708147].
\bibitem{Gibbons:1997xz}
G.~W.~Gibbons,
Nucl.\ Phys.\ B {\bf 514}, 603 (1998)
[arXiv:hep-th/9709027].


\bibitem{Tseytlin:1997cs}
A.~A.~Tseytlin,
Nucl.\ Phys.\ B {\bf 501}, 41 (1997)
[arXiv:hep-th/9701125].

\bibitem{Myers:1999ps}
R.~C.~Myers,
JHEP {\bf 9912}, 022 (1999)
[arXiv:hep-th/9910053].


\bibitem{Constable:2001ag}
N.~R.~Constable, R.~C.~Myers and O.~Tafjord,
JHEP {\bf 0106}, 023 (2001)
[arXiv:hep-th/0102080].


\bibitem{Constable:2001kv}
N.~R.~Constable, R.~C.~Myers and O.~Tafjord,
arXiv:hep-th/0105035.


\bibitem{Karczmarek:2001pn}
J.~L.~Karczmarek and C.~G.~.~Callan,
JHEP {\bf 0205}, 038 (2002)
[arXiv:hep-th/0111133].

\bibitem{Constable:2002yn}
N.~R.~Constable and N.~D.~Lambert,
Phys.\ Rev.\ D {\bf 66}, 065016 (2002)
[arXiv:hep-th/0206243].


\bibitem{Cook:2003rx}
P.~Cook, R.~de Mello Koch and J.~Murugan,
Phys.\ Rev.\ D {\bf 68}, 126007 (2003)
[arXiv:hep-th/0306250].



\bibitem{Sutcliffe:1997ec}
P.~M.~Sutcliffe,
Int.\ J.\ Mod.\ Phys.\ A {\bf 12}, 4663 (1997)
[arXiv:hep-th/9707009].


\bibitem{Diaconescu:1996rk}
D.~E.~Diaconescu,
Nucl.\ Phys.\ B {\bf 503}, 220 (1997)
[arXiv:hep-th/9608163].


\bibitem{Tsimpis:1998zh}
D.~Tsimpis,
Phys.\ Lett.\ B {\bf 433}, 287 (1998)
[arXiv:hep-th/9804081].


\bibitem{Kapustin:1998pb}
A.~Kapustin and S.~Sethi,
Adv.\ Theor.\ Math.\ Phys.\  {\bf 2}, 571 (1998)
[arXiv:hep-th/9804027].

\bibitem{Chen:2002vb}
X.~g.~Chen and E.~J.~Weinberg,
Phys.\ Rev.\ D {\bf 67}, 065020 (2003)
[arXiv:hep-th/0212328].


\bibitem{Bachas:2000dx}
C.~Bachas, J.~Hoppe and B.~Pioline,
JHEP {\bf 0107}, 041 (2001)
[arXiv:hep-th/0007067].













\end{thebibliography}
\end{document}